\documentclass[pra,twocolumn,superscriptaddress,showpacs]{revtex4}

\usepackage{graphics}           
\usepackage{bm}                 
\usepackage{amsmath}            
\usepackage{amssymb}
\usepackage{verbatim}           
\usepackage{amsthm}             
\theoremstyle{plain}            

\def\bra#1{{\langle#1|}}
\def\ket#1{{|#1\rangle}}
\def\bracket#1#2{{\langle#1|#2\rangle}}

\begin{document}

\title{Quantum cellular automata and quantum field theory in two spatial dimensions}

\author{Todd A. \surname{Brun}}\email{tbrun@usc.edu}
\author{Leonard \surname{Mlodinow}}\email{lmlodinow@gmail.com}
\affiliation{Center for Quantum Information Science and Technology, University of Southern California, Los Angeles, California}

\date{\today}

\begin{abstract}
Quantum walks on lattices can give rise to one-particle relativistic wave equations in the long-wavelength limit. In going to multiple particles, quantum cellular automata (QCA) are natural generalizations of quantum walks. In one spatial dimension, the quantum walk can be ``promoted'' to a QCA that, in the long-wavelength limit, gives rise to the Dirac quantum field theory (QFT) for noninteracting fermions. This QCA/QFT correspondence has both theoretical and practical applications, but there are obstacles to similar constructions in two or more spatial dimensions. Here we show that a method of construction employing distinguishable particles confined to the completely antisymmetric subspace yields a QCA in two spatial dimensions that gives rise to the 2D Dirac QFT. Generalizing to 3D will entail some additional complications, but no conceptual barriers. We examine how this construction evades the ``no go'' results in earlier work.
\end{abstract}


\maketitle

\section{Introduction}

If one applies quantum theory to sets of bits, the system's Hamiltonian will cause them to evolve in time just as they do in a computer. It is natural, therefore, to view unitary evolution as a kind of information processing, and to investigate the connection between quantum systems and information. In the four decades since Paul Benioff hatched the idea of a quantum computer \cite{Benioff80}, the field of quantum information has grown into one of the pillars of theoretical physics. But how deep is the connection between information and the physical world? Does what we observe as the physical world rest upon a substrate of information processing? Is the universe itself an evolving set of bits, a quantum computer, and if so, what is its program?
 
One can trace the history of such questions to a famous 1982 talk in which Richard Feynman noted that classical computers are ill-suited to simulating quantum systems \cite{Feynman82}. The difficulties stem from the fact that the wavefunctions of a quantum system lie in a vector space whose dimension grows exponentially with the system size, and so even the task of recording a system’s state is an exponentially difficult problem on a classical machine. Could a quantum computer operating on qubits rather than bits avoid that daunting blowup, Feynman asked? Feynman made the ``guess'' that every finite quantum mechanical system can indeed be described by another system that at each point in space-time this system has only two possible base states (a qubit) corresponding to that point being either occupied or unoccupied. Today we know such a system as a quantum cellular automaton (QCA) \cite{Grossing88,Watrous91,Lloyd93,Fussy93,Watrous95,Durr96,Meyer96,Meyer96a,Meyer96b}.

In 1989 John Wheeler put his own spin on Feynman's vision in his famous ``it from bit'' talk in Tokyo, in which he stated: 

\begin{quote}
``Every it---every particle, every field of force even the space time continuum itself---derives its function, its meaning, its very existence entirely---even if in some contexts indirectly---from the apparatus-elicited answers to yes or no questions, binary choices, bits. It from bit symbolizes the idea that every item of the physical world has at bottom---at a very deep bottom, in most instances---an immaterial source and explanation; that what we call reality arises in the last analysis from the posing of yes-no questions and the registering of equipment-evoked responses; in short, that all things physical are information theoretic in origin\ldots'' ---John Wheeler \cite{Wheeler90}
\end{quote}

Wheeler’s speculation was vague, and though he is often quoted it was never clear exactly what he meant. Then, in 1996 Seth Lloyd proved that quantum computers could be programmed to simulate the behavior of arbitrary quantum systems whose dynamics are determined by local interactions \cite{Lloyd96}. That same year Bialynicki-Birula showed that a discrete time quantum walk (QW) could, in the low energy and continuous-time limit, yield the single-particle Dirac equation \cite{Bialynicki94}. (He called it a cellular automaton---a QCA---and even today the terms QW and QCA are sometimes employed interchangeably; but to be precise, a QW is the one-particle sector of a QCA \cite{AharonovY93,Ambainis01,AharonovD01,Kempe03}.)

Were Feynman and Wheeler correct in their vision? Are quantum field theories (QFTs) such as QED and the standard model, and even quantum gravity, obtainable as limits of discrete space-time QCA theories? In the last decade or so there has been a flurry of papers investigating that question; an affirmative answer would be essentially a concrete realization of Wheeler's ``it from bit.’’

It is important to note that the goal of such studies is not to derive the QCA by mimicking the dynamics of the QFT, but rather to construct a QCA from a simple set of principles and symmetries, and show that we recover well-known Lorentz invariant QFTs in the limit of continuous time, and at energies low enough that the lattice spacing (say, the Planck length) is not probed. Much of the recent work has been on the one-particle sector (QWs) \cite{Strauch06,Bracken07,Chandrashekar10,Chandrashekar11,Chandrashekar13,DAriano14,Arrighi14,Farrelly14,DAriano15,Bisio15,Bisio15b,Bisio15c,Bisio16,Mallick16,DAriano17,Raynal17,DAriano17b,MlodinowBrun18,Manighalam19,Maeda20}, but progress has been made towards multiparticle QCAs that yield QFTs in the long-wavelength limit \cite{Yepez13,Yepez16,Arrighi20,MlodinowBrun20}.

Being immune to ultraviolet divergences, QCAs may be of use in regularizing the corresponding QFT, and some authors emphasize their potential as algorithms for simulating the dynamics of QFTs on a quantum computer \cite{Farrelly19}. The QCA/QFT correspondence could also provide a novel viewpoint that will give new insight into QFTs. In the 1970s the realization that low-energy strong force interactions could not be treated perturbatively led to lattice field theories as an alternative for numerical calculations, but also as a source of new understanding---QFTs are now often defined as the limit of lattice theories \cite{Creutz83}. In that regard, QCAs have some advantages over lattice theories \cite{FarrellyThesis}. For one, if we want to discretize space and retain causality, then we must also discretize time. Lattice quantum field theories with local Hamiltonians in continuous time are not truly causal, while QCAs are. In addition, there is evidence that QCAs do not suffer from the plague of fermion doubling (extra low energy modes) \cite{Nielsen81,DeGrand06,Tong12}. Finally, it is interesting to note that one cannot, a priori, rule out the possibility that it is the QCA and not its corresponding QFT that is fundamental. With neutron interferometer technology not far beyond what exists today, it will be possible to detect the existence of the QCA spatial lattice structure even if its scale is that of the Planck length \cite{BrunMlodinow19}.

Though there has been a great deal of progress in understanding the QCA/QFT correspondence, almost all of it has had to do with QCAs and QFTs in one space dimension. The difficulty in going to higher dimensions is related to the requirement that a QCA must be local, which in the context of QCAs means that the transformation at each time step depends only on the states of ``nearby'' qubits. That constrains the theory in many respects, in particular with regard to the issue of {\it fermionization}. If one has a set of qubits representing the presence or absence of particles in some internal state at a given lattice point, the corresponding creation and annihilation operators will anticommute while the operators associated with different lattice points will commute.

One can remedy this problem by employing the Jordan-Wigner transformation \cite{Jordan28} to produce operators that anticommute at different points; but, under a rather general set of conditions, in greater than one dimension a Jordan-Wigner transformation will result in a non-local theory, i.e., the transformed theory will not be a QCA. This result about QCAs in more than one spatial dimension was proven for a very large class of models, \cite{MlodinowBrun20}. Alternatively, one could switch to a momentum space picture and define fermionic creation and annihilation operators for those states, then employ the local unitary QCA evolution (defined on the coordinate lattice) to determine how they transform in each time step. This leads to a complicated time evolution defined by sets of integral equations that lack a simple local interpretation.

In this paper we get around these issues by embedding the physical states of the theory in a larger Hilbert space, somewhat analogous to the manner in which the physical states of quantum theory are rays rather than vectors in the Hilbert space in which they are defined. More precisely, we define our theory as pertaining to {\it distinguishable} particles, and later restrict ourselves to the antisymmetric subspace as the physical space. We show that this theory of distinguishable particles is a QCA (albeit one with very high-dimensional local subsystems), and has truly local evolution. We show that this antisymmetrized subspace is preserved by the QCA evolution, that it is possible to define creation and annihilation operators that evolve simply and obey the usual anticommutation relations, and that in the long wavelength limit this yields the multiparticle Dirac QFT.  We demonstrate this method first in one spatial dimension, and then show that it can be straightforwardly extended to two spatial dimensions, evading the no-go theorem in \cite{MlodinowBrun20}.

\section{A QCA for distinguishable particles}
\label{sec:distinguishable}

\subsection{Quantum walks}

A quantum walk is a unitary analogue of a classical random walk.  In both cases, the ``walker'' is a particle that can reside at any one of a set of positions, labeled by $x$.  At each time step, the particle can move to a neighboring position.  (Here we assume that time is discrete.  We do not consider continuous-time random walks and quantum walks in this paper.)

We can think of these positions as being the vertices of a graph, with edges between neighboring vertices.  In a random walk, the particle has some probability to move to the neighboring vertices.  In a quantum walk, the evolution is unitary, and the particle can move to a superposition of positions.  For the quantum walks in this paper, the graph is always {\it regular}, with every vertex having the same number of neighbors.  The number of neighbors is the {\it degree} of the graph, labeled by $d$.  We can label each of the $d$ outgoing edges from a vertex $1,\ldots,d$, with that label describing a direction the particle can move.  To start with, we will assume that this graph is finite, but we will consider the infinite limit eventually.

To maintain both unitarity and nontrivial dynamics, the particle generally must have an internal degree of freedom, or ``coin space.''  The Hilbert space has the form $\mathcal{H}_{\rm QW} = \mathcal{H}_X \otimes \mathcal{H}_C$, where $\mathcal{H}_X$ is the Hilbert space of the particle position, and $\mathcal{H}_C$ is the space of the internal degree of freedom.  In a standard quantum walk, the evolution from one time to the next is given by a unitary evolution operator $U_{\rm QW}$ of the form
\begin{equation}
\ket{\psi_{t+\Delta t}} = U_{\rm QW}\ket{\psi_t} = \left( I\otimes C \right) \left( \sum_{j=1}^d S_j \otimes P_j \right) \ket{\psi_t} ,
\label{eq:QWalkEvolution}
\end{equation}
where the $\{S_j\}$ are shift operators that move the particle from its current position to its neighbor in the direction $j$.  The $\{P_j\}$ are a set of $d$ orthogonal projectors on the internal space; and $C$ is a unitary that acts on the internal space, often called the ``coin flip'' unitary. Unlike classical random walks, the evolution is invertible and interference plays an important role in the dynamics.

The idea is that the walk proceeds by a process analogous to a series of coin flips.  The projectors $\{P_j\}$ correspond to different faces of the coin, which indicate which direction to move; the unitary $C$ scrambles the faces, so that one does not constantly move in the same direction.   In the walk in one dimension, there are two projectors $\{P_\pm\}$ corresponding to shifts to the right or left along the line.  When we go to the walk on a two-dimensional lattice we will use a generalization of this standard form involving successive shifts along the different axes.

\subsection{Multiparticle walks}
\label{sec:multiparticle}

We want to go to the multiparticle case.  A simple way to describe a fixed number $N$ of particles is to have $N$ independent copies of the quantum walk:  $\mathcal{H}_N = \left( \mathcal{H}_{\rm QW} \right)^{\otimes N} = \mathcal{H}_{\rm QW} \otimes \cdots \otimes \mathcal{H}_{\rm QW}$.  The evolution operator is $U_N = \left(U_{\rm QW}\right)^{\otimes N} = U_{\rm QW} \otimes \cdots \otimes U_{\rm QW}$.  We can see that in this construction, the $N$ particles all evolve independently; we can think of this as a theory of $N$ distinguishable, non-interacting particles.  Each particle is labeled by which Hilbert space describes it.

We can go a step beyond this simple construction by describing a space that can include any number of particles up to some maximum $N_{\rm max}$.  The Hilbert space is
\begin{equation}
\mathcal{H}_{\rm total} = \mathcal{H}^{(1)} \otimes \mathcal{H}^{(2)} \otimes \cdots \otimes \mathcal{H}^{(N_{\rm max})} ,
\end{equation}
where $\mathcal{H}^{(j)}$ is the Hilbert space of the particle labeled $j$.  Since we would like to allow any number of particles, each particle $j$ may or may not be present.  We include this possibility by using the Hilbert space
\begin{equation}
\mathcal{H}^{(j)} = \mathcal{H}_{\rm QW} \oplus \mathrm{span}(\ket\omega) ,
\end{equation}
where $\ket\omega$ is a vaccum state indicating that particle $j$ is not present.  We extend the evolution operator to leave this vacuum state invariant:
\begin{equation}
U = U_{\rm QW} + \ket\omega\bra\omega .
\end{equation}
The overall evolution operator of the full space is
\begin{equation}
U_{\rm total} = U^{\otimes N_{\rm max}} .
\end{equation}

\subsection{Embedding in a cellular automaton}

This is still a theory of distinguishable, noninteracting particles, but we have allowed the particle number to vary.  The question then arises:  is this a QCA?  Yes---the theory above can be embedded in a manifestly local QCA.

To define this QCA, we define a Hilbert space
\begin{equation}
\mathcal{H}_{\rm QCA} = \bigotimes_x \mathcal{H}_x , \ \ \ \ \mathcal{H}_x = \bigotimes_{e\in [1,\ldots,d]} \mathcal{H}_{x,e} ,
\label{eq:QCAspace}
\end{equation}
where the local subsystem labeled by position $x$ and internal state $e$ has the Hilbert space
\begin{equation}
\mathcal{H}_{x,e} = \bigotimes_{t\in [1,\ldots,N_{\rm max}]} \mathcal{H}_{x,e,t} ,\ \ \ \ 
\mathcal{H}_{x,e,t} = \mathbb{C}^2 .
\end{equation}
Here the index $t$ labels the ``type'' of the particle.  The two basis states of $\mathcal{H}_{x,e,t}$ correspond to the fact that we may or may not have a particle of type $t$ at the point $x$ in the internal state $e$.  Now instead of all particles being distinguishable, we have $N_{\rm max}$ particle types, where in principle one can have multiple particles of type $t$ with different values of $x$ and $e$.  Each particle type $t$ essentially forms a parallel QCA.  These QCAs coexist without interacting.  The QCA dynamics for each particle type is chosen so that the one-particle sector matches the dynamics of the quantum  walk:
\begin{equation}
U_{\rm QCA} = \bigotimes_{t\in [1,\ldots,N_{\rm max}]} U_t ,\ \ \ \ U_t = \hat{C}_t \hat{\Sigma}_t ,
\end{equation}
for each particle type $t$.  The QCA shift operator $\hat{\Sigma}$ acts on the basis states of the QCA by moving the content 0 or 1 of each position $x$ and internal state $e$ to the neighboring position in the direction $e$ from $x$, with the same internal state $e$.

The QCA coin-flip operator $\hat{C}$ is a tensor product of operators acting at each position $x$,
\begin{equation}
\hat{C}_t = \bigotimes_x C_{x,t} ,
\end{equation}
where $C_{x,t}$ is a unitary operator acting on the Hilbert space
\begin{equation}
\mathcal{H}_{x,t} = \bigotimes_{e\in [1,\ldots,d]} \mathbb{C}^2 .
\end{equation}
There is some freedom in defining $C_{x,t}$, but it must satisfy these requirements:
\begin{enumerate}
\item $C_{x,t}$ is unitary;
\item $C_{x,t}$ conserves the number of particles (i.e., number of $\ket1$ states);
\item $C_{x,t}$ acts as the identity on the state $\ket0^{\otimes d}$;
\item $C_{x,t}$ acts on the single-particle subspace just like the unitary operator $C$ in the quantum walk in Eq.~(\ref{eq:QWalkEvolution}).
\end{enumerate}
With these requirements the single-particle states of the QCA of type $t$ will evolve exactly like the quantum walk in Eq.~(\ref{eq:QWalkEvolution}). 

Moreover, the dynamics thus defined are manifestly local.  The QCA system is a tensor product of local subsystems, as defined in Eq.~(\ref{eq:QCAspace}).  The unitary evolution is the product for two unitaries, $\hat{C}$ and $\hat\Sigma$, each of which is manifestly local:
\begin{equation}
U_{\rm QCA} = \hat{C} \hat{\Sigma} = \left(\bigotimes_t \hat{C}_t \right) \left(\bigotimes_t \hat\Sigma_t \right) .
\end{equation}
One difference, however, from most QCA definitions is that the local subsystem Hilbert spaces $\mathcal{H}_{x}$ tend to have very high dimensions:  $\mathrm{dim} \mathcal{H}_x = 2^{dN_{\rm max}}$.  This high dimensionality is one factor that enables the model defined in this paper to evade the no-go theorem proven in Ref.~\cite{MlodinowBrun20}.

If we restrict our initial state in $\mathcal{H}_{\rm QCA}$ to include only 0 or 1 particle of each type $t$, this will be exactly the same as the multiparticle model described in Sec.~\ref{sec:multiparticle} above.  The subspace of all states with only 0 or 1 particles of each type is exactly isomorphic to $\mathcal{H}_{\rm total}$ defined above, and by construction the QCA evolution operator $U_{\rm QCA}$ acts on this subspace in exactly the same way that $U_{\rm total}$ acts on $\mathcal{H}_{\rm total}$.  So we can think of $\mathcal{H}_{\rm total}$ as a subspace of $\mathcal{H}_{\rm QCA}$.

\subsection{The physical subspace}

Having defined our system of distinguishable particles as a subspace $\mathcal{H}_{\rm total}$ of the full QCA Hilbert space $\mathcal{H}_{\rm QCA}$, we can now construct a theory that resides in a restricted subspace of $\mathcal{H}_{\rm total}$.  We can decompose $\mathcal{H}_{\rm total}$ into the direct sum of subspaces containing exactly $n$ particles.  Let $(t_1,t_2,\ldots,t_n)$ be a set of $n$ distinct particle types $t_j \in [1,\ldots,N_{\rm max}]$.  Then we can define a subspace $\mathcal{H}^{(t_1,t_2,\ldots,t_n)}$ as the subspace of $\mathcal{H}_{\rm total}$ comprising all states with exactly $n$ particles of types $(t_1,t_2,\ldots,t_n)$.  To avoid double counting, we will adopt a convention $t_1 < t_2 < \cdots < t_n$.  Then
\begin{eqnarray}
\mathcal{H}_{\rm total} &=& \mathcal{H}_0 \oplus \left( \bigoplus_{(t_1)} \mathcal{H}^{(t_1)}\right) \oplus \left( \bigoplus_{(t_1,t_2)} \mathcal{H}^{(t_1,t_2)} \right) \oplus \cdots \nonumber\\
&& \oplus \left( \bigoplus_{(t_1,t_2,\ldots,t_n)} \mathcal{H}^{(t_1,t_2,\ldots,t_n)} \right) \oplus \cdots ,
\end{eqnarray}
where $\mathcal{H}_0$ is the one-dimensional space with no particles,
\[
\mathcal{H}_0 = \mathrm{span}(\ket\Omega) , \ \ \ \ \ket\Omega = \ket\omega^{N_{\rm max}} .
\]

Define the antisymmetric subspace $\mathcal{A}^{(t_1,t_2,\ldots,t_n)}$ to be the set of all states $\psi(x_1,e_1; x_2,e_2; \cdots; x_n,e_n)$ in $\mathcal{H}^{(t_1,t_2,\ldots,t_n)}$ such that, if we interchange any two particles, the state is multiplied by $-1$.  That is, if $\pi$ is a permutation of the labels $[1,\ldots,n]$, and $p(\pi)$ is the parity of $\pi$ ($p(\pi) = 0$ if $\pi$ is even and $1$ if $\pi$ is odd), then
\begin{eqnarray}
&& \psi(x_{\pi(1)},e_{\pi(1)}; x_{\pi(2)},e_{\pi(2)}; \cdots; x_{\pi(n)},e_{\pi(n)}) \nonumber\\
&& = (-1)^{p(\pi)} \psi(x_1,e_1; x_2,e_2; \cdots; x_n,e_n) .
\end{eqnarray}

We now make the following observations:
\begin{enumerate}
\item The evolution $U_{\rm total}$ conserves the particle number of each type.
\item The evolution $U_{\rm total}$ maps the antisymmetric subspace $\mathcal{A}^{(t_1,t_2,\ldots,t_n)}$ to itself.
\item All $n$-particle subspaces are exactly isomorphic and have identical dynamics.  That is, if we have an $n$-particle state with particles of types $t_1,t_2,\ldots,t_n$, it will evolve exactly the same way as the analogous state with a different set of particle types $t_1',\ldots,t_n'$.
\end{enumerate}
This means that if we want to describe a state of $n$ particles, we can without loss of generality just use the first $n$ particle types, $t_1 = 1$, $t_2 = 2$, etc.  Based on these observations, we define the ``physical subspace'' $\mathcal{H}_{\rm phys}$ to be
\begin{equation}
\mathcal{H}_{\rm phys} = \mathcal{H}_0 \oplus \mathcal{A}^{(1)} \oplus \mathcal{A}^{(1,2)} \oplus \cdots \oplus \mathcal{A}^{(1,\ldots,N_{\rm max})} .
\end{equation}
If the initial state of the system is in the physical subspace, then as it evolves under $U_{\rm total}$ it will remain in the physical subspace at all times.

\section{The 1D QCA}

Let us now apply this construction to the 1D QCA.  This is based on the one-dimensional quantum walk,  This acts on a Hilbert space $\mathcal{H}_{\rm pos} \otimes \mathcal{H}_{\rm coin}$, where the position space is spanned by basis states $\ket{x}$ where $x=j\Delta x$, and the coin space has basis states $\ket{R}$ and $\ket{L}$.  To keep everything simple at first, we will assume that the position space is finite with $N$ locations corresponding to $j=0,1,\ldots,N-1$ and periodic boundary conditions, $\ket{N\Delta x} \equiv \ket{0}$.  The evolution operator is
\begin{equation}
U_{\rm 1D} = \left( I\otimes C \right) \left( S\otimes\ket{R}\bra{R} + S^\dagger\otimes\ket{L}\bra{L} \right) ,
\label{eq:1Devolution}
\end{equation}
where
\[
S\ket{x} = \ket{x+\Delta x} ,
\]
and $C$ is a unitary that acts on the coin space.  The exact form of $C$ is not that important; we will assume that
\begin{equation}
C = e^{i\theta Q} ,\ \ \ \ Q\ket{R} = \ket{L},\ \ \ \ Q\ket{L} = \ket{R} ,
\end{equation}
where  $Q=Q^\dagger$ and $Q^2 = I$. (We can always make such a choice, up to a global phase.)

From this quantum walk we can build a QCA following the procedure outlined in Sec.~\ref{sec:distinguishable} above.  We will now construct a basis for the physical space of this QCA, and show that it can be mapped straightforwardly onto a theory of free fermions in one dimension in the long-wavelength limit.

\subsection{Energy eigenstates of the 1D quantum walk}

The first step of this construction is to find the eigenstates of the unitary evolution operator in Eq.~(\ref{eq:1Devolution}).  We begin by defining a set of momentum basis states:
\begin{equation}
\ket{k} = \frac{1}{\sqrt{N}} \sum_{j=0}^N e^{-ikj\Delta x} \ket{x} ,\ \ \ \ k = 2\pi\ell/(N\Delta x) ,
\end{equation}
where $\ell = -N/2+1, \ldots, -1, 0, 1, \ldots, N/2$.  The momentum $k$ also has periodic boundary conditions, with $k + 2\pi m/\Delta x \equiv k$ for any integer $m$.  These states $\ket{k}$ are eigenstates of the shift operators:
\begin{equation}
S\ket{k} = e^{ik/\Delta x} \ket{k} ,\ \ \ S^\dagger\ket{k} = e^{-ik/\Delta x} \ket{k} .
\end{equation}
It is easy to check that
\begin{eqnarray}
U_{\rm 1D} \ket{k}\otimes\ket{R} &=& e^{ik\Delta x}\ket{k}\otimes \left( \cos(\theta) \ket{R} + i\sin(\theta) \ket{L} \right) , \nonumber\\
U_{\rm 1D} \ket{k}\otimes\ket{L} &=& e^{-ik\Delta x}\ket{k} \left( \cos(\theta) \ket{L} + i\sin(\theta) \ket{R} \right) .
\end{eqnarray}

We see that these momentum states are not eigenstates of $U_{\rm 1D}$, but they decompose the Hilbert space into two-dimensional subspaces, $\mathrm{span}(\ket{k}\otimes\ket{R},\ket{k}\otimes\ket{L})$, that are preserved by $U_{\rm 1D}$.  A vector $\alpha\ket{k}\otimes\ket{R} + \beta\ket{k}\otimes\ket{L}$ transforms under $U_{\rm 1D}$ as
\begin{eqnarray}
\left( \begin{array}{c} \alpha \\ \beta \end{array} \right) &\rightarrow& \mathbf{M} \left( \begin{array}{c} \alpha \\ \beta \end{array} \right) \\
&&\equiv \left(\begin{array}{cc} e^{ik\Delta x} \cos(\theta) & i e^{-ik\Delta x} \sin(\theta) \\
i e^{ik\Delta x} \sin(\theta) & e^{-ik\Delta x} \cos(\theta) \end{array} \right)
\left( \begin{array}{c} \alpha \\ \beta \end{array} \right) .\nonumber
\label{eq:transformMatrix}
\end{eqnarray}
It can also be convenient to write $\mathbf{M}$ in terms of the Pauli matrices:
\begin{eqnarray}
\mathbf{M} &=& \cos(k\Delta x)\cos(\theta) I + i \cos(k\Delta x)\sin(\theta) \sigma_X \\
&& + i \sin(k\Delta x)\sin(\theta) \sigma_Y + i \sin(k\Delta x)\cos(\theta) \sigma_Z . \nonumber
\end{eqnarray}
We can diagonalize this matrix and find the eigenvalues of $U_{\rm 1D}$:
\begin{eqnarray}
\lambda_{k,\pm} &\equiv& e^{\pm i \phi_k} \\
&=& \cos(k\Delta x)\cos(\theta) \pm i \sqrt{1-\cos^2(k\Delta x) \cos^2(\theta)} , \nonumber
\end{eqnarray}
with eigenvectors
\begin{equation}
\mathbf{v}_{k,\pm} = \frac{1}{\mathcal{N}_\pm} \left( \begin{array}{c}
\sin(k\Delta x)\cos(\theta) \pm \sqrt{1-\cos^2(k\Delta x) \cos^2(\theta)} \\
e^{ik\Delta x}\sin(\theta) \end{array} \right) ,
\label{eq:QWeigenvectors}
\end{equation}
with normalizations
\begin{equation}
\mathcal{N}_\pm = \sqrt{ \begin{array}{c} \biggl(\sin(k\Delta x)\cos(\theta) \\
\pm \sqrt{1-\cos^2(k\Delta x) \cos^2(\theta)}\biggr)^2 \\
+ \sin^2(\theta) \end{array}} .
\label{eq:normalizations}
\end{equation}
The eigenstates of $U_{\rm 1D}$ are therefore
\begin{equation}
\ket{k,\pm} \equiv \left(\begin{array}{c} \ket{k}\otimes\ket{R} \\ \ket{k}\otimes\ket{L} \end{array}\right) \cdot \mathbf{v}_{k,\pm} ,
\end{equation}
and have eigenvalues $\lambda_{k,\pm} = e^{\pm i\phi_k}$.  The phases of the eigenvalues are $\pm\phi_k$, where $\cos(\phi_k) = \cos(k\Delta x)\cos(\theta)$, or $\phi_k = \cos^{-1}(\cos(k\Delta x)\cos(\theta))$.

\subsection{The energy basis of the 1D QCA}

We can use the energy eigenstates of the 1D quantum walk to construct a basis (which we will call the {\it energy basis}) for the 1D QCA space $\mathcal{H}_{\rm phys}$.  These are antisymmetric combinations of tensor products of the energy eigenstates defined above.  Let the number of particles present be $n$.  For $n=0$ there is just one state, the vacuum:
\begin{equation}
\ket\Omega_{\rm phys} \equiv \ket\omega^{\otimes N_{\rm max}} .
\end{equation}
For $n=1$ we have one particle (of type 1) in an energy eigenstate:
\begin{equation}
\ket{k,\pm}_{\rm phys} \equiv \ket{k,\pm}\otimes \ket\omega^{\otimes N_{\rm max}-1} .
\end{equation}
For $n=2$ we have two particles (of types 1 and 2) in energy eigenstates:
\begin{widetext}
\begin{equation}
\ket{k_1,\varepsilon_1; k_2,\varepsilon_2}_{\rm phys} \equiv \frac{1}{\sqrt2}
\left(\ket{k_1,\varepsilon_1}\otimes \ket{k_2,\varepsilon_2} -
\ket{k_2,\varepsilon_2}\otimes \ket{k_1,\varepsilon_1} \right) \otimes
\ket\omega^{\otimes N_{\rm max}-2} ,
\end{equation}
where $\varepsilon_{1,2} = \pm$.  And for arbitrary $n$,
\begin{equation}
\ket{k_1,\varepsilon_1; \cdots; k_n,\varepsilon_n}_{\rm phys} \equiv \frac{1}{\sqrt{n!}} \sum_\pi
(-1)^{p(\pi)} \ket{k_{\pi(1)},\varepsilon_{\pi(1)}}\otimes \cdots \otimes
\ket{k_{\pi(n)},\varepsilon_{\pi(n)}} \otimes \ket\omega^{\otimes N_{\rm max}-n} ,
\label{eq:antisymmetrizedStates}
\end{equation}
\end{widetext}
where $\pi$ is a permutation of $[1,\ldots,n]$ and $p(\pi)$ is the parity 1 or 0 (for odd or even) of the permutation $\pi$.

These antisymmetrized states $\ket{k_1,\varepsilon_1; \cdots; k_n,\varepsilon_n}_{\rm phys}$ for all values of $n$ from 0 to $N_{\rm max}$ form a basis for the physical subspace $\mathcal{H}_{\rm phys}$.  Moreover, they are all eigenstates of the evolution operator $U_{\rm QCA}$ with eigenvalues
\begin{equation}
\lambda_{k_1,\varepsilon_1; \cdots; k_n,\varepsilon_n} = e^{i\sum_{j=1}^n \varepsilon_j \phi_{k_j}} .
\end{equation}
There is one minor ambiguity, however, that must be removed.  As defined above, for example,
$\ket{k_1,\varepsilon_1; k_n,\varepsilon_2}_{\rm phys} = - \ket{k_2,\varepsilon_2; k_1,\varepsilon_1}_{\rm phys}$.  So if we let the $k_j$'s take arbitrary values then the basis would be overcomplete.  We can remove this ambiguity by establishing a fixed ordering on the pairs $k,\varepsilon$, and requiring that the labels $k_1,\varepsilon_1; \cdots; k_n,\varepsilon_n$ be listed in the correct order.  It does not matter what ordering we choose.  For this 1D case, a convenient ordering would be to list $k_1 \le k_2 \le \cdots \le k_n$; and if both $k,+$ and $k,-$ are present, to list $k,-$ first. But any choice will work equally well.  Note also that these pairs $k,\varepsilon$ cannot be repeated; the antisymmetrization would then yield zero.  So $k_1,\varepsilon_1; \cdots; k_n,\varepsilon_n$ must all be distinct.

\subsection{Creation and annihilation operators}

Having defined a set of basis states for $\mathcal{H}_{\rm phys}$ in Eq.~(\ref{eq:antisymmetrizedStates}), we can formally define a set of creation and annihilation operators that move us between these states.  We can define the creation and annihilation operators by their actions on the energy basis states of $\mathcal{H}_{\rm phys}$:
\begin{equation}
\ket{k_1,\varepsilon_1; \cdots; k_n,\varepsilon_n}_{\rm phys} = a^\dagger_{k_1,\varepsilon_1} \cdots a^\dagger_{k_n,\varepsilon_n} \ket\Omega_{\rm phys} ,
\end{equation}
\[
a_{k,\varepsilon}\ket\Omega_{\rm phys} = 0 ,
\]
where these operators obey the usual anticommutation relations:
\begin{eqnarray}
\{a_{k_1,\varepsilon_1}, a_{k_2,\varepsilon_2}\} &=& \{a^\dagger_{k_1,\varepsilon_1}, a^\dagger_{k_2,\varepsilon_2}\} = 0 ,\nonumber\\
\{a^\dagger_{k_1,\varepsilon_1}, a_{k_2,\varepsilon_2}\} &=& \delta_{k_1 k_2} \delta_{\varepsilon_1 \varepsilon_2} I .
\end{eqnarray}

Given that we have assumed a maximum number of distinct particle types $N_{\rm max}$, we must also have the somewhat unusual property
\begin{equation}
a^\dagger_{k,\varepsilon} \ket{k_1,\varepsilon_1; \cdots; k_{N_{\rm max}},\varepsilon_{N_{\rm max}}}_{\rm phys} = 0\, \forall k,\varepsilon .
\label{eq:maxParticles}
\end{equation}
We can avoid having this extra condition, however, if we choose $N_{\rm max} = 2N$, where $N$ is the total number of lattice sites; in that case, the condition in Eq.~(\ref{eq:maxParticles}) is automatically satisfied by the usual property of fermionic creation operators $\left(a^\dagger_{k,\varepsilon}\right)^2 = 0$.  For the purposes of this model, we will make this assumption, and then allow $N\rightarrow\infty$.

As defined, these operators are clearly nonlocal; but that is not a problem, since the underlying dynamics of the QCA {\it are} local.  We have also not defined how these operators act on states outside of $\mathcal{H}_{\rm phys}$, but again, this doesn't really matter.  We can choose, for example, to have them annihilate all states orthogonal to $\mathcal{H}_{\rm phys}$.  How do these operators evolve under the unitary $U_{\rm 1D}$?  We can see that
\begin{eqnarray}
&& U_{\rm 1D} \ket{k_1,\varepsilon_1; \cdots; k_n,\varepsilon_n}_{\rm phys} \nonumber\\
&=& U_{\rm 1D} a^\dagger_{k_1,\varepsilon_1} \cdots a^\dagger_{k_n,\varepsilon_n} \ket\Omega_{\rm phys} \nonumber\\
&=& U_{\rm 1D} a^\dagger_{k_1,\varepsilon_1} U^\dagger_{\rm 1D} U_{\rm 1D} \cdots U^\dagger_{\rm 1D} U_{\rm 1D} a^\dagger_{k_n,\varepsilon_n} U^\dagger_{\rm 1D} U_{\rm 1D} \ket\Omega_{\rm phys} \nonumber\\
&=& \left(U_{\rm 1D} a^\dagger_{k_1,\varepsilon_1} U^\dagger_{\rm 1D}\right)  \cdots \left(U_{\rm 1D} a^\dagger_{k_n,\varepsilon_n} U^\dagger_{\rm 1D}\right) \ket\Omega_{\rm phys} \nonumber\\
&=& e^{i\sum_{j=1}^n \varepsilon_j \phi_{k_j}} \ket{k_1,\varepsilon_1; \cdots; k_n,\varepsilon_n}_{\rm phys} ,
\end{eqnarray}
which implies that
\begin{equation}
\left(U_{\rm 1D} a^\dagger_{k,\varepsilon} U^\dagger_{\rm 1D}\right) = e^{i\varepsilon \phi_k} a^\dagger_{k,\varepsilon}  .
\end{equation}
This very simple time evolution, and the relationship between the creation operators and the basis states, allows us to write the effective evolution operator $U_{\rm 1D}$ on the physical subspace $\mathcal{H}_{\rm phys}$.  That is:
\begin{equation}
U_{\rm 1D} = \exp\left\{ i \sum_k \phi_k \left( a^\dagger_{k,+} a_{k,+} - a^\dagger_{-k,-} a_{-k,-}\right) \right\} .
\end{equation}

We can also define a momentum representation, and write $U_{\rm 1D}$ in terms of that. By diagonalizing the matrix $\mathbf{M}$ in Eq.~(\ref{eq:transformMatrix}) we defined the energy basis states for the quantum walk that in turn were used to define the energy basis states of the QCA, and the creation and annihilation operators $a^\dagger_{k,\pm}$, $a_{k,\pm}$ that transform between them.  By going back to the original basis we can define creation operators $a^\dagger_{k,R}$, $a^\dagger_{k,L}$ and their corresponding annihilation operators for momentum basis states.  We write
\begin{eqnarray}
\left(\begin{array}{c} 1 \\ 0 \end{array}\right) &=& \alpha_R \mathbf{v}_{k,+} + \beta_R \mathbf{v}_{k,-} ,\nonumber\\
\left(\begin{array}{c} 0 \\ 1 \end{array}\right) &=& \alpha_L \mathbf{v}_{k,+} + \beta_L \mathbf{v}_{k,-} ,
\end{eqnarray}
where $\mathbf{v}_{k,\pm}$ are given in Eq.~(\ref{eq:QWeigenvectors}), and solve for the coefficients
\begin{equation}
\left(\begin{array}{c} \alpha_R \\ \beta_R \end{array}\right)
= \frac{1}{\mathcal{N}_R} \left(\begin{array}{c} \mathcal{N}_+ \\ -\mathcal{N}_- \end{array}\right) ,
\end{equation}
\begin{widetext}
\begin{equation}
\left(\begin{array}{c} \alpha_L \\ \beta_L \end{array}\right)
= \frac{e^{-ik\Delta x}}{\mathcal{N}_L} \left(\begin{array}{c} \mathcal{N}_+ \left(- \sin(k\Delta x)\cos(\theta) + \sqrt{1-\cos^2(k\Delta x) \cos^2(\theta)} \right) \\ \mathcal{N}_- \left(\sin(k\Delta x)\cos(\theta) + \sqrt{1-\cos^2(k\Delta x) \cos^2(\theta)} \right) \end{array}\right) ,
\end{equation}
\end{widetext}
where
\begin{eqnarray}
\mathcal{N}_R &=& 2 \sqrt{1-\cos^2(k\Delta x) \cos^2(\theta)} , \nonumber\\
\mathcal{N}_L &=& 2 \sin(\theta) \sqrt{1-\cos^2(k\Delta x) \cos^2(\theta)} ,
\end{eqnarray}
and $\mathcal{N}_\pm$ are defined in Eq.~(\ref{eq:normalizations}). The creation operators for momentum states are
\begin{eqnarray}
a^\dagger_{k,R} &=& \alpha_R a^\dagger_{k,+} + \beta_R a^\dagger_{k,-} , \nonumber\\
a^\dagger_{k,L} &=& \alpha_L a^\dagger_{k,+} + \beta_L a^\dagger_{k,-} .
\end{eqnarray}
By construction, these evolve by
\begin{eqnarray}
U_{\rm 1D} a^\dagger_{k,R} U^\dagger_{\rm 1D} &=& e^{ik\Delta x}\cos(\theta) a^\dagger_{k,R} \nonumber\\
&& + ie^{-ik\Delta x}\sin(\theta) a^\dagger_{k,L} , \nonumber\\
U_{\rm 1D} a^\dagger_{k,L} U^\dagger_{\rm 1D} &=& ie^{ik\Delta x}\sin(\theta) a^\dagger_{k,R}\nonumber\\
&& + e^{-ik\Delta x}\cos(\theta) a^\dagger_{k,L} .
\label{eq:momentumRep}
\end{eqnarray}
From the momentum representation defined above we can derive a Dirac field theory in the long-wavelength limit.

\subsection{The Dirac Equation in 1D}

Define a long wavelength limit with $|k\Delta x|,|\theta| \ll 1$.  Let the duration of one time step be defined as $\Delta t$, and define the ``speed of light'' to be $c \equiv \Delta x/\Delta t$ and the ``rest mass'' to be $mc^2 \equiv \hbar\theta/\Delta t$.  Expanding the matrix elements in Eq.~(\ref{eq:momentumRep}) to linear order, we see that
\begin{eqnarray}
\left(\begin{array}{c} U_{\rm 1D} a^\dagger_{k,R} U^\dagger_{\rm 1D} \\
U_{\rm 1D} a^\dagger_{k,L} U^\dagger_{\rm 1D} \end{array}\right) &\approx&
\left(\begin{array}{c} a^\dagger_{k,R} \\ a^\dagger_{k,L} \end{array}\right) +
ik\Delta x \left(\begin{array}{c} a^\dagger_{k,R} \\ - a^\dagger_{k,L} \end{array}\right) \nonumber\\
&& + i\theta \left(\begin{array}{c} a^\dagger_{k,L} \\ a^\dagger_{k,R} \end{array}\right) .
\label{eq:longWavelength1D}
\end{eqnarray}
We can define a ``time derivative'' superoperator
\begin{equation}
\partial_t O = (1/\Delta t)\left( U_{\rm 1D} O U^\dagger_{\rm 1D} - O \right) ,
\end{equation}
and rewrite Eq.~(\ref{eq:longWavelength1D}) as
\begin{equation}
i\hbar \partial_t \left(\begin{array}{c} a^\dagger_{k,R} \\ a^\dagger_{k,L} \end{array}\right) \approx
\left( - pc \sigma_Z - mc^2 \sigma_X \right) \left(\begin{array}{c} a^\dagger_{k,R} \\ a^\dagger_{k,L} \end{array}\right) ,
\end{equation}
where the momentum is $p=\hbar k$. This is the evolution equation for the Dirac field in one dimension, where $\sigma_X$ and $\sigma_Z$ play the roles of $\gamma_0$ and $\gamma_0 \gamma_1$, respectively.

If we instead look at the energy representation, in the long-wavelength limit we can write the eigenvalues of $U_{\rm 1D}$ in the form
\begin{equation}
\lambda_{k,\pm} = e^{\pm i\phi_k} \equiv e^{\mp iE_k \Delta t/\hbar} ,
\end{equation}
and expanding $\phi_k$ to first order we get
\begin{eqnarray}
E_k \equiv \hbar\phi_k/\Delta t &\approx& (\hbar/\Delta t)\sqrt{ k^2\Delta x^2 + \theta^2}\nonumber\\
&=& \sqrt{ p^2 c^2 + m^2 c^4 } ,
\end{eqnarray}
which is the usual classical formula for the energy of a relativistic particle with rest mass $m$ and momentum $p$.

In the derivation presented here, the vacuum state is the state with no particles present. However, it is possible to introduce a ``Dirac sea'' construction, as described in \cite{MlodinowBrun20}, in which all negative energy states are occupied. In such a construction, antiparticles can be interpreted as holes in the space of negative energy states.

\section{The 2D QCA}

The same kind of procedure can be done for higher-dimensional quantum walks.  As shown in Ref.~\cite{MlodinowBrun18}, a quantum walk on the 2D square lattice can also be defined using a two-dimensional internal coin space, which yields the 2D Dirac equation in the long-wavelength limit.  We will demonstrate the construction above in this case, which overcomes the difficulties described in Ref.~\cite{MlodinowBrun20} and will allow us to recover the Dirac field theory in two spatial dimensions.  For the present we will assume that this lattice is finite, $N\times N$, with periodic boundary conditions (so it is equivalent to a torus), which gives a dimension $2N^2$ for the overall Hilbert space.  Later we will let $N\rightarrow\infty$.

The particular quantum walk on the 2D square lattice from \cite{MlodinowBrun18} has an evolution unitary that takes the form
\begin{eqnarray}
U_{\rm 2D} &=& \left( I\otimes C \right) \left( S_Y\otimes\ket{U}\bra{U} + S_Y^\dagger\otimes\ket{D}\bra{D} \right) \nonumber\\
&& \times \left( S_X\otimes\ket{R}\bra{R} + S_X^\dagger\otimes\ket{L}\bra{L} \right) ,
\label{eq:2Devolution}
\end{eqnarray}
where the coin-flip operator can again be written
\begin{equation}
C = e^{i\theta Q} .
\end{equation}
More on $Q$ in a moment, but first note that the 2D shift operators act as one would expect on the position of the particle,
\begin{eqnarray}
S_X\ket{x,y} = \ket{x+\Delta x,y} , && S_Y\ket{x,y} = \ket{x,y+\Delta x} , \nonumber\\
S_X^\dagger\ket{x,y} = \ket{x-\Delta x,y} , && S_Y^\dagger\ket{x,y} = \ket{x,y-\Delta x} .
\end{eqnarray}
We can use a vector notation $\ket{x,y} = \ket{\mathbf{x}}$, where $\mathbf{x}$ is the 2D real vector
\[
\mathbf{x} = \left(\begin{array}{c} x \\ y \end{array}\right) \equiv \left(\begin{array}{c} q\Delta x \\ r\Delta x \end{array}\right) ,
\]
where $q$ and $r$ are integers modulo $N$.

The projectors on the coin space correspond to a pair of unbiased bases:
\begin{eqnarray}
\bracket{R}{L} &=& \bracket{U}{D} = 0, \nonumber\\
| \bracket{R}{U} | &=& | \bracket{R}{D} | = | \bracket{L}{U} | =  | \bracket{L}{D} | \nonumber\\
&=& 1/\sqrt2 .
\end{eqnarray}
This last condition is equivalent to the ``equal norm condition'' in Ref.~\cite{MlodinowBrun18}.  As in the 1D case, $Q=Q^\dagger$ and $Q^2 = I$. $Q$ is chosen to switch the particle's direction, but now it must do so for both the $X$ and the $Y$ directions:
\begin{eqnarray}
Q \ket{R} = \ket{L} ,\ \ \ && Q\ket{L} = \ket{R} ,\nonumber\\
Q \ket{U} = \ket{D} ,\ \ \ && Q\ket{D} = \ket{U} .
\end{eqnarray}
As shown in Ref.~\cite{MlodinowBrun18}, this implies that if we define operators $\Delta P_X = \ket{R}\bra{R} - \ket{L}\bra{L}$ and $\Delta P_Y = \ket{U}\bra{U} - \ket{D}\bra{D}$, then the three operators $\Delta P_X$, $\Delta P_Y$ and $Q$ all mutually anticommute,
\begin{equation}
\{\Delta P_X, \Delta P_Y\} = \{\Delta P_X,Q\} = \{\Delta P_Y,Q\} = 0 ,
\end{equation}
and also
\begin{equation}
\left(\Delta P_X\right)^2 = \left(\Delta P_Y\right)^2 = Q^2 = I .
\end{equation}

\subsection{Energy eigenstates of the 2D quantum walk}

Just as in the 1D case, we begin by transforming to the momentum picture:
\begin{equation}
\ket{\mathbf{k}} = \frac{1}{N} \sum_{x,y} e^{-i\mathbf{k}\cdot\mathbf{x}} \ket{\mathbf{x}} ,
\end{equation}
where
\[
\mathbf{k} = \left(\begin{array}{c} k_X \\ k_Y \end{array}\right) \equiv \left(\begin{array}{c} 2\pi n/(N\Delta x) \\ 2\pi o/(N\Delta x) \end{array}\right) ,
\]
where $n$ and $o$ are integers modulo $N$.  It is easy to see that $\ket{\mathbf{k}}$ is an eigenstate of $S_X$ and $S_Y$:
\begin{equation}
S_X\ket{\mathbf{k}} = e^{i k_X\Delta x} \ket{\mathbf{k}} , \ \ \ 
S_Y\ket{\mathbf{k}} = e^{i k_Y\Delta x} \ket{\mathbf{k}} .
\end{equation}
We can rewrite the evolution operator in this momentum representation:
\begin{equation}
U_{\rm 2D} = e^{i\theta \left( I\otimes Q\right)} e^{i \left(K_Y\otimes \Delta P_Y\right)\Delta x} e^{i \left(K_X\otimes \Delta P_X\right)\Delta x},
\end{equation}
where $K_X$ and $K_Y$ are the operators corresponding to the $X$ and $Y$ components of the momentum:
\[
K_X \ket{\mathbf{k}} = k_X \ket{\mathbf{k}} , \ \ \ K_Y \ket{\mathbf{k}} = k_Y \ket{\mathbf{k}} .
\]
The two-dimensional subspace spanned by $\ket{\mathbf{k}} \otimes \ket{R}$ and $\ket{\mathbf{k}} \otimes \ket{L}$ will therefore be preserved by the evolution operator $U_{\rm 2D}$.

Let us make the specific choice $\Delta P_X = \sigma_X$, $\Delta P_Y = \sigma_Y$, and $Q = \sigma_Z$ (which is the same as assuming that $\{\ket{R},\ket{L}\}$ is the $\sigma_Z$ eigenbasis).  (Any other choice that meets the requirements above will give the same dynamics up to a change of basis for the coin space.)  Then the superposition vector $\alpha\ket{\mathbf{k}}\otimes\ket{R} + \beta\ket{\mathbf{k}}\otimes\ket{L}$ is transformed by $U_{\rm 2D}$ as
\begin{equation}
\left(\begin{array}{c} \alpha \\ \beta \end{array}\right) \rightarrow
\mathbf{M} \left(\begin{array}{c} \alpha \\ \beta \end{array}\right) ,
\end{equation}
where the $2\times2$ matrix $\mathbf{M}$ can be written
\begin{eqnarray}
\mathbf{M} &=& \bigl( \cos(k_X\Delta x) \cos(k_Y\Delta x) \cos(\theta) \nonumber\\
&& - \sin(k_X\Delta x) \sin(k_Y\Delta x) \sin(\theta) \bigr) I \nonumber\\
&& + i \bigl( \cos(k_X\Delta x) \cos(k_Y\Delta x) \sin(\theta) \nonumber\\
&& + \sin(k_X\Delta x) \sin(k_Y\Delta x) \cos(\theta) \bigr) \sigma_X \nonumber\\
&& + i \bigl( - \cos(k_X\Delta x) \sin(k_Y\Delta x) \cos(\theta) \nonumber\\
&&+ \sin(k_X\Delta x) \cos(k_Y\Delta x) \sin(\theta) \bigr) \sigma_Y \nonumber\\
&& + i \bigl( \sin(k_X\Delta x) \cos(k_Y\Delta x) \cos(\theta) \nonumber\\
&& + \cos(k_X\Delta x) \sin(k_Y\Delta x) \sin(\theta) \bigr) \sigma_Z .
\label{eq:2DMatrix}
\end{eqnarray}

The matrix $\mathbf{M}$ is a $2\times2$ unitary of the form
\[
\mathbf{M} = r_0 I + i\left( r_1 \sigma_X + r_2 \sigma_Y + r_3 \sigma_Z \right) ,
\]
where $r_0^2 + r_1^2 + r_2^2 + r_3^2 = 1$; its eigenvalues are
\begin{equation}
\lambda_{\pm} \equiv e^{\pm i \phi} = \cos(\phi) \pm i\sin(\phi) ,
\label{eq:2Deigenvalues}
\end{equation}
where $\cos(\phi) = r_0$ and $\sin{\phi} = \sqrt{r_1^2 + r_2^2 + r_3^2}$, and the eigenvectors take the form
\begin{eqnarray}
\mathbf{v}_+ &=& \frac{1}{\sqrt2} \left(\begin{array}{c} \frac{\sin(\phi)+r_3}{\sqrt{\sin^2(\phi) + r_3\sin(\phi)}} \\
\frac{r_1 + ir_2}{\sqrt{\sin^2(\phi) + r_3\sin(\phi)}} \end{array} \right) ,\nonumber\\
\mathbf{v}_- &=& \frac{1}{\sqrt2} \left(\begin{array}{c} - \frac{\sin(\phi)-r_3}{\sqrt{\sin^2(\phi) - r_3\sin(\phi)}} \\
\frac{r_1 + ir_2}{\sqrt{\sin^2(\phi) - r_3\sin(\phi)}} \end{array} \right) .
\label{eq:2Deigenvectors}
\end{eqnarray}

Plugging in the values from the matrix in Eq.~(\ref{eq:2DMatrix}) we have
\begin{eqnarray}
r_0 &=& \cos(k_X\Delta x) \cos(k_Y\Delta x) \cos(\theta) \nonumber\\
&& - \sin(k_X\Delta x) \sin(k_Y\Delta x) \sin(\theta) , \nonumber\\
r_1 &=& \cos(k_X\Delta x) \cos(k_Y\Delta x) \sin(\theta) \nonumber\\
&& + \sin(k_X\Delta x) \sin(k_Y\Delta x) \cos(\theta) , \nonumber\\
r_2 &=& - \cos(k_X\Delta x) \sin(k_Y\Delta x) \cos(\theta) \nonumber\\
&&+ \sin(k_X\Delta x) \cos(k_Y\Delta x) \sin(\theta) , \nonumber\\
r_3 &=& \sin(k_X\Delta x) \cos(k_Y\Delta x) \cos(\theta) \nonumber\\
&& + \cos(k_X\Delta x) \sin(k_Y\Delta x) \sin(\theta) .
\end{eqnarray}
The solution in Eqs.~(\ref{eq:2Deigenvalues}--\ref{eq:2Deigenvectors}) tells us that the energy eigenstates of the 2D quantum walk have eigenvalues
\begin{equation}
\lambda_{\mathbf{k},\pm} \equiv e^{\pm i \phi_{\mathbf{k}}} ,
\end{equation}
where $\phi_{\mathbf{k}} = \cos^{-1}(r_0)$.  We can label the energy eigenstates $\ket{\mathbf{k},\pm}$, where
\begin{equation}
\ket{\mathbf{k},\pm} = \alpha_\pm \ket{\mathbf{k}}\otimes\ket{R} + \beta_\pm \ket{\mathbf{k}}\otimes\ket{L} ,
\end{equation}
and the coefficients $\alpha_\pm$ and $\beta_\pm$ are taken from the eigenvectors $\mathbf{v}_\pm$ in Eq.~(\ref{eq:2Deigenvectors}).

\subsection{Creation and annihilation operators}

Just as we did in the 1D case, we can now define a set of basis states for the subspace $\mathcal{H}_{\rm phys}$ in two spatial dimensions.  For $n=0$ particles there is a unique vacuum state:
\begin{equation}
\ket\Omega_{\rm phys} \equiv \ket\omega^{\otimes N_{\rm max}} .
\end{equation}
For $n=1$ we have one particle (of type 1) in an energy eigenstate:
\begin{equation}
\ket{\mathbf{k},\pm}_{\rm phys} \equiv \ket{\mathbf{k},\pm}\otimes \ket\omega^{\otimes N_{\rm max}-1} .
\end{equation}
For $n=2$ we have two particles (of types 1 and 2) in energy eigenstates:
\begin{widetext}
\begin{equation}
\ket{\mathbf{k}_1,\varepsilon_1; \mathbf{k}_2,\varepsilon_2}_{\rm phys} \equiv \frac{1}{\sqrt2}
\left(\ket{\mathbf{k}_1,\varepsilon_1}\otimes \ket{\mathbf{k}_2,\varepsilon_2} -
\ket{\mathbf{k}_2,\varepsilon_2}\otimes \ket{\mathbf{k}_1,\varepsilon_1} \right) \otimes
\ket\omega^{\otimes N_{\rm max}-2} ,
\end{equation}
where $\varepsilon_{1,2} = \pm$.  And for arbitrary $n$,
\begin{equation}
\ket{\mathbf{k}_1,\varepsilon_1; \cdots; \mathbf{k}_n,\varepsilon_n}_{\rm phys} \equiv \frac{1}{\sqrt{n!}} \sum_\pi
(-1)^{p(\pi)} \ket{\mathbf{k}_{\pi(1)},\varepsilon_{\pi(1)}}\otimes \cdots \otimes
\ket{\mathbf{k}_{\pi(n)},\varepsilon_{\pi(n)}} \otimes \ket\omega^{\otimes N_{\rm max}-n} ,
\end{equation}
\end{widetext}
where once again $\pi$ is a permutation of $[1,\ldots,n]$ and $p(\pi)$ is the parity 1 or 0 (for odd or even) of the permutation $\pi$.

Just as in the 1D case, we can remove the ambiguity by adopting an ordering convention on the pairs of indices $\mathbf{k},\varepsilon$, and requiring that $(\mathbf{k}_1,\varepsilon_1) < \cdots < (\mathbf{k}_n,\varepsilon_n)$.  However, unlike the 1D case, there is no choice of ordering that is obviously more natural than any other.  In fact, it does not matter to the theory which ordering is chosen.

As an example, to show that it is possible to choose a consistent ordering, we could first order the momentum vectors $\mathbf{k} = (k_x, k_y)$ where $\mathbf{k}_1 < \mathbf{k}_2$ if $(k_y)_1 < (k_y)_2$, or if $(k_y)_1 = (k_y)_2$ and $(k_x)_1 < (k_x)_2$.  If both $\mathbf{k},+$ and $\mathbf{k},-$ are present we list the $-$ state first.

As before, having defined a set of basis states for $\mathcal{H}_{\rm phys}$ we can formally define a set of creation and annihilation operators that move us between these states.  We can define the creation and annihilation operators by their actions on the energy basis states of $\mathcal{H}_{\rm phys}$:
\begin{equation}
\ket{\mathbf{k}_1,\varepsilon_1; \cdots; \mathbf{k}_n,\varepsilon_n}_{\rm phys} = a^\dagger_{\mathbf{k}_1,\varepsilon_1} \cdots a^\dagger_{\mathbf{k}_n,\varepsilon_n} \ket\Omega_{\rm phys} ,
\end{equation}
\[
a_{\mathbf{k},\varepsilon}\ket\Omega_{\rm phys} = 0 ,
\]
where these operators obey the usual anticommutation relations:
\begin{eqnarray*}
\{a_{\mathbf{k}_1,\varepsilon_1}, a_{\mathbf{k}_2,\varepsilon_2}\} &=& \{a^\dagger_{\mathbf{k}_1,\varepsilon_1}, a^\dagger_{\mathbf{k}_2,\varepsilon_2}\} = 0 ,\\
\{a^\dagger_{\mathbf{k}_1,\varepsilon_1}, a_{\mathbf{k}_2,\varepsilon_2}\} &=& \delta_{\mathbf{k}_1 \mathbf{k}_2} \delta_{\varepsilon_1 \varepsilon_2} I .
\end{eqnarray*}
Again, if a creation operator $a_{\mathbf{k},\pm}$ acts an a state that already has the maximum number of particles $N_{\rm max}$ it must annihilate that state.  We can get rid of this problem by having $N_{\rm max}$ equal the maximum number of available sites $2N^2$.

From the way that the basis states are defined in terms of energy eigenstates, we can easily see that the creation operators transform under the time evolution operator by acquiring a phase:
\begin{equation}
\left(U_{\rm 2D} a^\dagger_{\mathbf{k},\varepsilon} U^\dagger_{\rm 2D}\right) = e^{i\varepsilon \phi_{\mathbf{k}}} a^\dagger_{\mathbf{k},\varepsilon} .
\end{equation}
Just as in the 1D case, we can use this to define creation operators for momentum states, by inverting the transformation to the eigenvector basis in Eq.~(\ref{eq:2Deigenvectors}):
\begin{equation}
\left(\begin{array}{c} 1 \\ 0 \end{array}\right) = \alpha_R \mathbf{v}_+ + \beta_R \mathbf{v}_- ,
\end{equation}
which has solutions
\begin{eqnarray}
\alpha_R &=& \sqrt{\frac{\sin(\phi)+r_3}{2\sin(\phi)}} ,\nonumber\\
\beta_R &=& \sqrt{\frac{\sin(\phi)-r_3}{2\sin(\phi)}} , \nonumber\\
\alpha_L &=& \frac{r_1 - i r_2}{\sqrt{2\sin(\phi)(\sin(\phi)+r_3)}} ,\nonumber\\
\beta_L &=& \frac{r_1 - i r_2}{\sqrt{2\sin(\phi)(\sin(\phi)-r_3)}} .
\end{eqnarray}
We can use this to define creation operators for momentum states:
\begin{eqnarray}
a^\dagger_{\mathbf{k},R} &=& \alpha_R a^\dagger_{\mathbf{k},+} + \beta_R a^\dagger_{\mathbf{k},-} ,\nonumber\\
a^\dagger_{\mathbf{k},L} &=& \alpha_L a^\dagger_{\mathbf{k},+} + \beta_L a^\dagger_{\mathbf{k},-} .
\end{eqnarray}
These have time evolution
\begin{equation}
\left(\begin{array}{c} a^\dagger_{\mathbf{k},R} \\ a^\dagger_{\mathbf{k},R} \end{array}\right) 
\rightarrow \left(\begin{array}{c} U_{\rm 2D} a^\dagger_{\mathbf{k},R} U^\dagger_{\rm 2D} \\ U_{\rm 2D} a^\dagger_{\mathbf{k},R} U^\dagger_{\rm 2D} \end{array}\right)
= \mathbf{M} \left(\begin{array}{c} a^\dagger_{\mathbf{k},R} \\ a^\dagger_{\mathbf{k},R} \end{array}\right) ,
\label{eq:momentumEvol2D}
\end{equation}
where the matrix $\mathbf{M}$ is given in Eq.~(\ref{eq:2DMatrix}).

\subsection{The Dirac Equation in 2D}

Going again to the limit where $|\mathbf{k}|\Delta x \ll 1$ and $|\theta| \ll 1$, we can recover the 2D Dirac field in the long-wavelength limit.  Considering the time evolution in the momentum description given by Eq.~(\ref{eq:momentumEvol2D}), we can approximate
\begin{eqnarray}
\left(\begin{array}{c} U_{\rm 2D} a^\dagger_{\mathbf{k},R} U^\dagger_{\rm 2D} \\
U_{\rm 2D} a^\dagger_{\mathbf{k},L} U^\dagger_{\rm 2D} \end{array}\right) &\approx&
\left(\begin{array}{c} a^\dagger_{\mathbf{k},R} \\ a^\dagger_{\mathbf{k},L} \end{array}\right) +
ik_X\Delta x \left(\begin{array}{c} a^\dagger_{\mathbf{k},R} \\ - a^\dagger_{\mathbf{k},L} \end{array}\right) \nonumber\\
&& + ik_Y\Delta x \left(\begin{array}{c} i a^\dagger_{\mathbf{k},L} \\ -i a^\dagger_{\mathbf{k},R} \end{array}\right) \nonumber\\
&& + i\theta \left(\begin{array}{c} a^\dagger_{\mathbf{k},L} \\ a^\dagger_{\mathbf{k},R} \end{array}\right) . 
\label{eq:longWavelength2D}
\end{eqnarray}
Defining once again a ``time derivative'' superoperator
\begin{equation}
\partial_t O = (1/\Delta t)\left( U_{\rm 2D} O U^\dagger_{\rm 2D} - O \right) ,
\end{equation}
and making the same definitions $\mathbf{p} = \hbar \mathbf{k}$, $c = \Delta x/\Delta t$, and $mc^2 \equiv \hbar\theta/\Delta t$, Eq.~(\ref{eq:longWavelength2D}) becomes
\begin{equation}
i\hbar \partial_t \left(\begin{array}{c} a^\dagger_{\mathbf{k},R} \\ a^\dagger_{\mathbf{k},L} \end{array}\right) \approx
\left( - cp_X \sigma_Z + cp_Y \sigma_Y - mc^2 \sigma_X \right) \left(\begin{array}{c} a^\dagger_{\mathbf{k},R} \\ a^\dagger_{\mathbf{k},L} \end{array}\right) ,
\end{equation}
This is the evolution equation for the Dirac field in two dimension, where $\sigma_X$, $-\sigma_Y$ and $\sigma_Z$ play the roles of $\gamma_0$, $\gamma_0 \gamma_1$, and $\gamma_0\gamma_2$, respectively.

As for the energy eigenvalues, in the long-wavelength limit
\begin{equation}
\lambda_{k,\pm} = e^{\pm i\phi_k} \equiv e^{\mp iE_k \Delta t/\hbar} 
\end{equation}
becomes
\begin{equation}
E_k \equiv \hbar\phi_k/\Delta t \approx \sqrt{ p_X^2 c^2 + p_Y^2 c^2 + m^2 c^4 }
\end{equation}
to first order in $k_X\Delta x$, $k_Y\Delta x$ and $\theta$. This is the usual classical formula for the energy of a relativistic particle with rest mass $m$ and momentum $\mathbf{p}$. Just as in the 1D case described above, we can introduce a ``Dirac sea'' construction for the 2D QFT.

\section{Discussion and Future Work}

In this paper we have presented a cellular automaton construction that yields the Dirac field theory in one and two dimensions in the long-wavelength limit. We are quite confident that the same type of construction will work in three spatial dimensions as well. However, that will require a four-dimensional internal space, and is sufficiently more complicated that we defer it to a later publication.

This construction---based on confining a set of distinguishable particles to their completely antisymmetric subspace---evades the no-go result demonstrated in Ref.~\cite{MlodinowBrun20}, which rules out a much simpler family of QCA constructions in dimensions higher than 1. It does so at the cost of requiring a QCA whose local subsystems are very high dimensional, and in producing the theory only within an antisymmetric subspace of a larger Hilbert space. However, having introduced this more complex system, the time-evolution of the physical basis states becomes a simple phase rotation, which allows the definition of creation and annihilation operators with simple time evolution as well.

Much remains to be understood. The extension of this paper's construction to three dimensions is conceptually straightforward, but will entail some mathematical complications that go beyond those in this paper. A good QCA model of bosons will not suffer from the no-go result in \cite{MlodinowBrun20}; but how to combine bosons and fermions in a single QCA theory of interacting particles is not at all obvious. It will also be interesting to inquire whether such a theory, with an underlying discrete spacetime implying that Lorentz invariance is only approximately valid at high energies, would have observable consequences that could be tested experimentally, as in Ref.~\cite{BrunMlodinow19}. We plan to tackle these questions in our future work.

\begin{acknowledgments}

TAB acknowledges useful conversations with Namit Anand, Christopher Cantwell, Yi-Hsiang Chen, Shengshi Pang, Prithviraj Prabhu and Chris Sutherland.  LM would like to thank Erhard Seiler and Alois Kabelschacht of the Max-Planck-Institute in Munich for fruitful discussions.  The authors are grateful for the hospitality of John Preskill and Caltech's Institute for Quantum Information and Matter (IQIM).

\end{acknowledgments}


\end{document}